\documentclass[aps,twocolumn,preprintnumbers,nofootinbib,superscriptaddress]{revtex4}
\usepackage{amsmath, amssymb} \usepackage{graphicx} \usepackage{amsfonts}
\usepackage{array} \usepackage{amsthm} \usepackage{bm}
\usepackage{palatino} \usepackage{mathpazo} 
\usepackage{supertabular}

\usepackage[breaklinks]{hyperref}
\usepackage{color}

\newcommand{\nn}{\nonumber}
\newcommand{\be}{\begin{equation}}
\newcommand{\ee}{\end{equation}}
\newcommand{\ba}{\begin{eqnarray}}
\newcommand{\ea}{\end{eqnarray}}
\newcommand{\bal}{\begin{align}}
\newcommand{\eal}{\end{align}}

\newcommand{\e}{{\rm e}}
\newcommand{\dd}{{\rm d}}

\newcommand{\bb}{\bibitem}

\newcommand{\bw}{\begin{widetext}}
\newcommand{\ew}{\end{widetext}}

\begin{document}
\title{Gaussian and Weibull noncommutative charged black holes}

\author{Mustapha Azreg-A\"{\i}nou}
\affiliation{Ba\c{s}kent University, Engineering Faculty, Ba\u{g}l\i ca Campus, Ankara 06810, Turkey}


\begin{abstract}
We derive and investigate the physical properties of asymptotically flat noncommutative regular charged black holes with a Gaussian mass density distribution and a Weibull electric charge density distribution. Both distributions replace the Dirac one and introduce a set of substitution rules constituting new ways of counting. The solutions have a de Sitter behavior in the vicinity of the origin provided the electric charge density is Weibull of the form $r^{n/2}{\rm e}^{-r^2/(4\theta^2)}$ with $n\geq 1$. The electric field and temperature are finite for all values of the radial coordinate, mass, and charge. The charge is bounded from above for stability reasons and stable charged quantum particles have mass and charge bounded from below and from above within the simplified semi-classical model we present in this work.
\end{abstract}


\maketitle

\section{How should the metric of a noncommutative black hole be?\label{seci}}
Some classes of regular black holes~\cite{regular1}-\cite{cons}, known in the literature, are manifestations of nonlinear electrodynamics or manifestations of nonlocal, extended, finite mass and charge densities. Other classes include matched de Sitter-Schwarzschild solutions.

Constructing finite mass (energy) and charge distributions seems to be an endless process unless there are physical or/and geometrical arguments supporting it; different distributions lead by \emph{integration of the field equations} to different metric regular solutions.

A first real tentative to justify the construction of a finite mass distribution~\cite{SS1} emerged in the noncommutative geometry of spacetime, which introduces an anti-symmetric real matrix $\theta^{\mu\nu}$ proportional to the commutator $[\mathbf{x}^{\mu},\mathbf{x}^{\nu}]$. Under the assumption that $\theta^{\mu\nu}$ being diagonal, with nonvanishing components equal to $\pm \theta^2$ where $\theta$ is a constant having a dimension of length, it has been shown that the use of expectation values of operators between coherent states results in a Gaussian distribution---instead of a Dirac distribution---of the scalar product of two mean positions~\cite{SS1}. The resulting Gaussian distribution has a width $\theta$ and reduces to a Dirac distribution in the limit $\theta\to0$.

Inspired by such a result, it was argued in Ref.~\cite{SS2} that the most important effects of noncommutativity, curing the pathological short distance behaviors, are incorporated in the theory of general relativity upon using extended Gaussian mass distributions of the form\footnote{This idea has been applied in Ref.~\cite{comment} to derive noncommutative charged black holes.}
\begin{equation}\label{i1}
    \rho_m(r)=\frac{M}{(2\pi)^{3/2}\theta^3}~\e^{-r^2/(2\theta^2)},
\end{equation}
where $M$ is the total mass of the regular Schwarzschild black hole
\begin{equation}\label{i2}
    M=\int_0^{\infty}\rho_m(r)4\pi r^2\dd r.
\end{equation}
The rhs of~\eqref{i1} replaces the mass Dirac distribution, $M\delta(r)/(2\pi r^2)$, which is known to lead to the singular Schwarzschild black hole. Here the width $\theta$ represents the limit beyond which classical arguments apply and it is a measure of the uncertainty in the value of $r$: $\Delta r\geq \theta$. For $r<\theta$ semi-classical or quantum approaches are needed to describe the physics of the black hole.

If $m(r)$ denotes the mass inside a sphere of radius $r$,
\begin{equation}\label{i3}
m(r)=\int_0^{r}\rho_m(r')4\pi r'^2\dd r'=\frac{2M}{\sqrt{\pi}}~\int_0^{r^2/(2\theta^2)} \sqrt{t}\e^{-t}\dd t,
\end{equation}
[where we have set $t=r'^2/(2\theta^2)$] the metric of the regular Schwarzschild black hole is given in terms of the lower and upper incomplete gamma functions~\cite{nist},
\begin{align*}
&\gamma(\alpha,z)\equiv \int_0^{z}t^{\alpha-1}\e^{-t}\dd t,
&\Gamma(\alpha,z)\equiv \int_z^{\infty}t^{\alpha-1}\e^{-t}\dd t,
\end{align*}
[where $\gamma(\alpha,z)+\Gamma(\alpha,z)=\Gamma(z)$ if $\alpha\neq 0,\,-1,\,-2,\,\dots$ with $\Gamma(z)$ being the Euler gamma function] by
\begin{align}\label{i4}
&\dd s^2=g(r)\dd t^2-\frac{\dd r^2}{g(r)}-r^2(\dd\vartheta^2+\sin^2\vartheta \dd\varphi^2),\\
&g(r)=1-\frac{4M}{r\sqrt{\pi}}\gamma \big(\frac{3}{2},\frac{r^2}{2 \theta ^2}\big)=1-\frac{2M\gamma \big(\frac{3}{2},\frac{r^2}{2 \theta ^2}\big)}{r\Gamma \big(\frac{3}{2}\big)}.\nn
\end{align}
This regular solution reduces to the singular Schwarzschild black hole in the limit $\theta\to0$:
\[\lim_{\theta\to0}g(r)=1-2M/r.\]

Mathematically speaking, the limits $\theta\to0$ and $r/\theta\to\infty$, carried on the previous expression and on the subsequent expressions encountered in this paper, are the same and yield the same output; however, their physical meanings are different. The former limit implies that if the effects of noncommutativity are small or suppressed one recovers the singular classical solutions for all values $r$. By the latter limit it is understood that the effects of noncommutativity are neglected far away from the source. This is but another way to state that the effects of noncommutativity are more important for distances of the order of the Compton wavelength, which are inversely proportional to $M$.

Depending on the values of $M$, the metric~\eqref{i4} describes a quantum particle\footnote{We are using the notation of~\cite{SS3} where we replace $a$ by $\theta$. This is the same as using the notation of~\cite{SS2} where we replace $4\theta$ by $2\theta^2$.} ($M<M_0$), an extremal black hole ($M=M_0$), or a nonextremal black hole with two horizons ($M>M_0$)~\cite{SS2,SS3}, where $M_0\equiv 1.9\theta/\sqrt{2}$. The metric is regular everywhere including the origin where it has the de Sitter behavior as it should be,
\begin{equation}\label{i5}
    g\underset{(r\to 0)}{\simeq}1-\sqrt{\frac{2}{\pi}}~\frac{2M}{3\theta^3}~r^2.
\end{equation}
The de Sitter behavior in the vicinity of the origin ensures a vacuum negative pressure resisting the collapse of matter into a singularity. This is a well-known statement since the works of Sakharov and Gliner~\cite{Gliner1,Gliner2}, in that, the core of collapsing matter, with high matter density, should have a cosmological-type equation of state $\epsilon=-p_r>0$, where $\epsilon$ is the energy density and $p_r$ is the radial pressure.

Note that the solution~\eqref{i4} has been derived on choosing the stress-energy tensor (SET), $T^{\mu\nu}$, to be that of an anisotropic perfect fluid~\cite{SS2,SS3} with only nonvanishing diagonal components: an energy-density $\epsilon=\rho_m$~\eqref{i1}, a radial pressure $p_r=-\rho_m$, and a tangential pressure $p_t=-\rho_m-r\partial_r\rho_m/2$. Both pressures ($p_r,\,p_t$) are negative. However, ordinary matter cannot have negative pressure, so the source is in fact made of two perfect fluids: A \emph{pressureless} dust with $\epsilon_{\text{d}}=\rho_m$ and a quantum vacuum having $\epsilon_{\text{v}}=0$ and negative pressures $p_r=-\rho_m<0$ and $p_t=-\rho_m-r\partial_r\rho_m/2<0$:
\begin{align}\label{i6}
&\text{dust: }& &\epsilon_{\text{d}}=\rho_m,\; \text{ no pressures};\\
&\text{vacuum: }& &\epsilon_{\text{v}}=0,\;p_r=-\rho_m,\;p_t=-\rho_m-\tfrac{r\partial_r\rho_m}{2}.\nn
\end{align}
By the method of superposition of fields and its generalization~\cite{tol}-\cite{2f}, using the same reference frame one can add the components of the two fluids to have, for instance, $\epsilon_{\text{total}}=\epsilon_{\text{d}}+\epsilon_{\text{v}}=\rho_m$. This is the way by which one includes the effects of noncommutativity translating the ``substitution rule"~\cite{SS2}: Dirac distribution to Gaussian distribution\footnote{In spherical coordinates we have, \[\int_0^{\infty}\frac{\delta(r)}{2\pi r^2}~4\pi r^2\dd r=\int_0^{\infty}\frac{\e^{-r^2/(2\theta^2)}}{(2\pi)^{3/2}\theta^3}~4\pi r^2\dd r=1.\]},
\begin{equation}\label{i7}
\frac{\delta(r)}{2\pi r^2}\to \frac{\e^{-r^2/(2\theta^2)}}{(2\pi)^{3/2}\theta^3},
\end{equation}
via the introduction of a mass density for pressureless ordinary matter and vacuum negative pressures.

Vacuum fluctuations may lead to $\epsilon_{\text{v}}=e>0$ where $e$ is some positive contribution that is generally function of position. Since $\rho_m$, attributable to collapsing matter, is assumed to be large in the vicinity of the origin, which is the only region where $e$ may have observable effects, one may safely ignore the effects of $e$.

It is well-known that the consideration of quantum effects leads to the violation~\cite{anec} of some of the local energy conditions (LEC). From~\eqref{i6} we have $\rho_m+p_r+2p_t<0$ which violates the strong energy condition; the other LECs (null, weak, and dominant energy conditions) are not violated.

If the noncommutative regular black hole were charged, how would its charge be distributed? Does the substitution rule Dirac-to-Gauss~\eqref{i7} still hold? The purpose of this work is to answer these questions by constructing a noncommutative charged black hole. Quantum mechanically there are no arguments to support the construction of such black holes, we will stick to our approach~\cite{cons} by which we already constructed charged cores extending neutral cores derived in Ref.~\cite{IrinaD}. We will impose the constraint that the noncommutative regular black hole be of the de Sitter type in the vicinity of the origin and modify the substitution rule Dirac-to-Gauss~\eqref{i7} for charge distributions. Our approach constitutes an alternative to, and shares some common features with, that used to derive the charged solution of Ref.~\cite{comment}.

Aren't there other observable effects of noncommutativity other than negative pressures that sustain matter from collapsing? We will obtain the results that (a) for a given mass $M$ the ratio $Q^2/M^2$ at the extremality condition, where $Q$ is the electric charge, is smaller for a noncommutative regular black hole than it is for its Reissner-Nordstr\"om counterpart, (b) for a given $r$ and $Q$ the electric field is always less than that of a Reissner-Nordstr\"om black hole, (c) the electric field and temperature are finite for all $r$, (d) a stability condition sets an upper bound for the charge, and (e) stable charged quantum particles have mass and charge obeying $1.346\theta<M<4.254\theta$ and $2.394\theta<|Q|<4.255\theta$.

\begin{figure}
\centering
\includegraphics[width=0.43\textwidth]{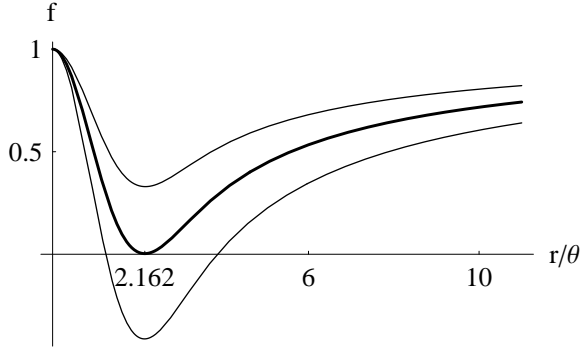} \\
\caption{\footnotesize{Plots of the metric $f$ versus $r/\theta$ for $Q^2G/(4 \pi  \epsilon _0c^4)=0.5$. Upper plot: A (non-black-hole) charged quantum particle~\eqref{w7} for $MG/c^2=1$. Intermediate plot: An extremal noncommutative charged black hole~\eqref{w7} for $MG/c^2=1.44$. Lower plot: A noncommutative charged black hole~\eqref{w7} with two horizons for $MG/c^2=2$.}}\label{Fig1}
\end{figure}

\section{Dirac versus Gauss versus Weibull distributions\label{secdgw}}
We seek a solution of the form
\begin{align}\label{w1}
&\dd s^2=f(r)\dd t^2-\frac{\dd r^2}{f(r)}-r^2(\dd\vartheta^2+\sin^2\vartheta \dd\varphi^2),\\
&f(r)=1-\frac{2m(r)}{r}+\frac{q^2(r)}{r^2},\nn
\end{align}
where $m(r)$ is given by~\eqref{i3} and
\begin{equation}\label{w2}
q^2(r)=\int_0^{r}\rho_{q^2}(r')4\pi r'^2\dd r'.
\end{equation}
Here $\rho_{q^2}(r)$ is a measure of the local charge density squared. Note that by~\eqref{i5} the term $2m(r)/r\propto r^2$ in the limit $r\to 0$. We want that the term $q^2(r)/r^2$ behaves in the same way so that $f$ will have a de Sitter behavior in that limit. A simple way to realize that is to choose the electric charge density squared, $\rho_{q^2}(r)$, proportional to the mass density~\eqref{i1} times $r^n$ with $n\geq 1$. For simplicity we choose $n=1$ so that
\begin{equation}\label{w3}
\rho_{q^2}(r)\propto r\rho_m(r).
\end{equation}
The total charge squared inside a sphere of radius $r$~\eqref{w2} takes the form
\begin{multline}\label{w4}
q^2(r)=4\pi C\int_0^{r} r'^3\e^{-r'^2/(2\theta^2)}\dd r'\\=8\pi \theta^4C\int_0^{r^2/(2\theta^2)} t\e^{-t}\dd t
=8\pi \theta^4C\gamma \Big(2,\frac{r^2}{2 \theta ^2}\Big),
\end{multline}
where $C$ is independent of $r$ [here again we have set $t=r'^2/(2\theta^2)$]. With $\int_0^{\infty} t\e^{-t}\dd t=\gamma(2,\infty)=1$, we obtain
\begin{equation}\label{w5}
    Q^2=8\pi \theta^4C,
\end{equation}
where $Q$ is the total charge of the black hole defined by $Q^2=\int_0^{\infty}\rho_{q^2}(r')4\pi r'^2\dd r'$. This implies
\begin{equation}\label{w6}
\rho_{q^2}(r)=\frac{Q^2r\e^{-r^2/(2\theta^2)}}{8\pi\theta^4}.
\end{equation}
This implies that the electric charge density is proportional to $\sqrt{r}\e^{-r^2/(4\theta^2)}$.

\subsection{noncommutative charged black hole - Charged quantum particle}
Finally, the metric component $f$ takes the form where we have introduced all physical constants:
\begin{equation}\label{w7}
f=1-\frac{2 M G}{c^2 r}~\frac{\gamma \big(\frac{3}{2},\frac{r^2}{2 \theta ^2}\big)}{\Gamma \big(\frac{3}{2}\big)}+\frac{Q^2 G}{4 \pi  \epsilon _0c^4 r^2}~\gamma \Big(2,\frac{r^2}{2 \theta ^2}\Big).
\end{equation}
This can be brought to the following form:
\begin{multline}\label{w7b}
\hspace{-1.9mm}f=1-\frac{2 M G}{c^2 r}+\frac{Q^2 G}{4 \pi  \epsilon _0c^4 r^2}\\
\hspace{-1.9mm}+\frac{2 M G}{c^2 r}~\frac{\Gamma \big(\frac{3}{2},\frac{r^2}{2 \theta ^2}\big)}{\Gamma \big(\frac{3}{2}\big)}-\frac{Q^2 G}{4 \pi  \epsilon _0c^4 r^2}\Big(1+\frac{r^2}{2 \theta ^2}\Big)\e^{-r^2/(2\theta^2)},
\end{multline}
where we have used $\gamma(3/2,z)+\Gamma(3/2,z)=\Gamma(3/2)$ and $\gamma(2,z)=1-(1+z)\e^{-z}$.

In our approach, instead of the substitution~\eqref{i7}, we have the following Dirac-to-Weibull substitution rule for an electric charge density squared:
\begin{equation}\label{i8}
\frac{\delta(r)}{2\pi r^2}\to \frac{r\e^{-r^2/(2\theta^2)}}{8\pi\theta^4},
\end{equation}

It is a matter of computer algebra systems to check that
\[\lim_{\theta\to 0}f=1-\frac{2 M G}{c^2 r}+\frac{Q^2 G}{4 \pi  \epsilon _0c^4 r^2},\]
which is the Reissner-Nordstr\"om black hole. Thus, reversing the order in the substitutions~\eqref{i7} and~\eqref{i8} yields the well-known solutions of classical general relativity.

Figure~\ref{Fig1}, which is a plot of $f$ versus $r/\theta$, depicts the three cases of a (non-black-hole) charged quantum particle, an extremal noncommutative charged black hole, and a noncommutative charged black hole for a fixed value of $Q^2G/(4 \pi  \epsilon _0c^4)=0.5$ and three different values of $MG/c^2=1,\,1.44,\,2$, respectively.

Note that the Weibull distribution~\eqref{w6} has a maximum value at $r=\theta$. The total charge inside the spherical shell of radius $r=\theta$, $\sqrt{1-3/(2\sqrt{\e})}~Q\simeq 0.3Q$, represents 30 per cent of the total charge $Q$. The electric push on the shell due to the 70 per cent of the charge outside the shell and the gravitational pull are balanced by the electric push due to the 30 per cent of the charge distributed fuzzily inside the shell and the vacuum negative radial pressure ensuring the equilibrium of the shell. This argument applies to the equilibrium of any shell of arbitrary radius $r$.

The metric $f$ behaves in the vicinity of the origin as
\begin{equation}\label{w8}
f\underset{(r\to 0)}{\simeq}1-\frac{\frac{16}{3}\sqrt{\frac{2}{\pi}}~\frac{MG}{c^2\theta}-\frac{Q^2 G}{4 \pi  \epsilon _0c^4 \theta^2}}{8\theta^2}~r^2,
\end{equation}
which reduces to~\eqref{i5} if $Q=0$. To have a de Sitter behavior near the origin, the charge of the black hole should be bounded from above by
\begin{equation}\label{w9}
\frac{Q^2 G}{4 \pi  \epsilon _0c^4 \theta^2} < \frac{16}{3}\sqrt{\frac{2}{\pi}}~\frac{MG}{c^2\theta}.
\end{equation}
We consider this as a condition of stability. this sets an upper bound for the charge. A common feature with the charged solution discussed in Ref.~\cite{comment} is that such a bound exists too. In fact, the solution (13) of Ref.~\cite{comment}, has a de Sitter behavior near the origin (using the notation of that reference),
\begin{equation*}
g_{00}\underset{(r\to 0)}{\simeq}1-\frac{4\sqrt{\pi\theta}~M-\sqrt{2}Q^2}{12\pi\theta^2}~r^2,
\end{equation*}
provided $\sqrt{2}Q^2<4\sqrt{\pi\theta}~M$.

Asymptotically ($r\to\infty$) the metric $f$~\eqref{w7b} approaches the Reissner-Nordstr\"om black hole solution,
\[f\underset{(r\to \infty, \,\theta >0)}{\simeq}1-\frac{2 M G}{c^2 r}+\frac{Q^2 G}{4 \pi  \epsilon _0c^4 r^2}+\cdots,\]
where the next dropped terms are proportional to $\e^{-r^2/(2\theta^2)}/r^{2n}$ with $n\in \mathbb{N}$.

\begin{figure}
\centering
\includegraphics[width=0.43\textwidth]{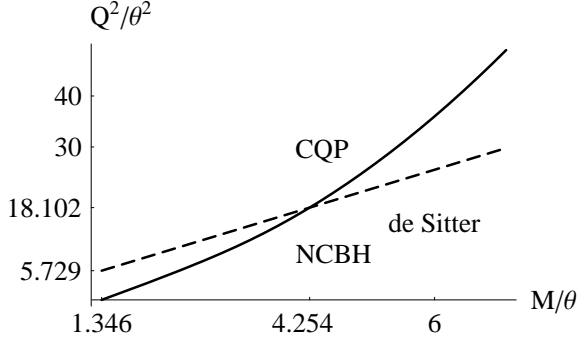} \\
\caption{\footnotesize{Plots of the extremality conditions~\eqref{w10}. The continuous curve represents the \emph{extremal} noncommutative charged black hole~\eqref{w7}, the region beneath it represents a noncommutative charged black hole (NCBH)~\eqref{w7}, and the region above it represents a (non-black-hole) charged quantum particle (CQP)~\eqref{w7}. The dashed straight line is a plot of~\eqref{w9} with the sign ``$<$" replaced by the sign ``$=$". The region beneath it satisfies the constraint~\eqref{w9} ensuring a de Sitter behavior in the vicinity of the origin.}}\label{Fig2}
\end{figure}
\begin{figure*}
\centering
\includegraphics[width=0.43\textwidth]{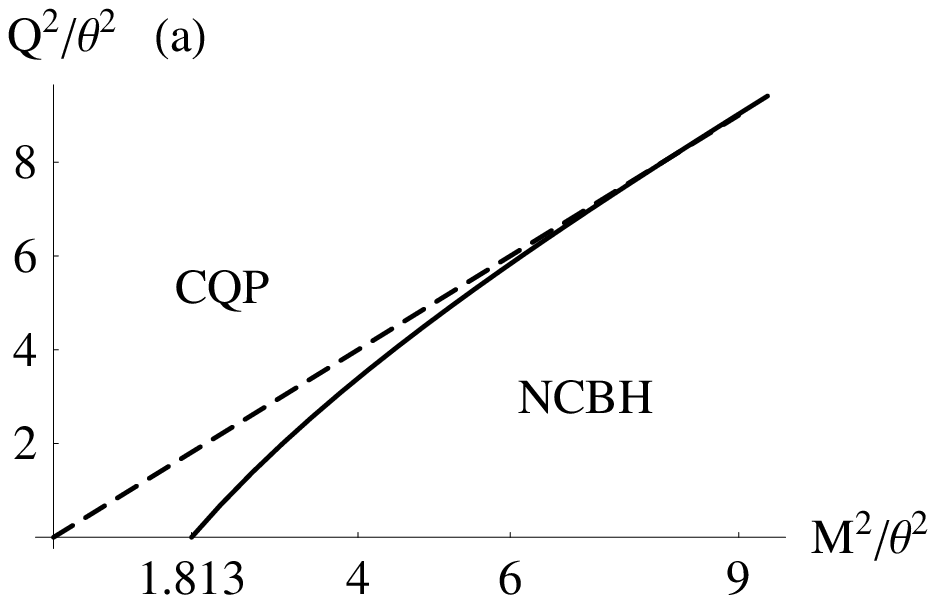}  \includegraphics[width=0.43\textwidth]{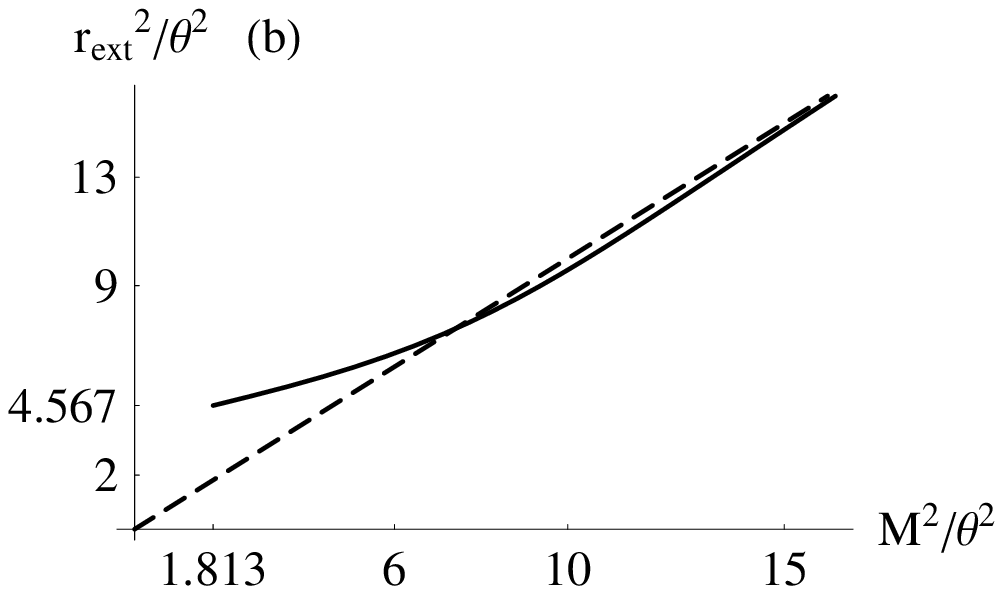} \\
\caption{\footnotesize{Other plots of the extremality conditions~\eqref{w10}. (a) $Q^2/\theta^2$ versus $M^2/\theta^2$ where we have made the substitutions $MG/c^2\to M$ and $Q^2G/(4 \pi  \epsilon _0c^4)\to Q^2$. The continuous curve represents the \emph{extremal} noncommutative charged black hole~\eqref{w7}, the region beneath it represents a noncommutative charged black hole (NCBH)~\eqref{w7}, and the region above it represents a (non-black-hole) charged quantum particle (CQP)~\eqref{w7}. The dashed curve represents the \emph{extremal} Reissner-Nordstr\"om black hole. For $Q=0$ the plots shows the relation $M_0=\sqrt{1.813}\,\theta=1.346\theta$ as found in~\cite{SS2}. (b) $r_{\text{ext}}^2/\theta^2$ versus $M^2/\theta^2$. The continuous curve represents the \emph{extremal} NCBH~\eqref{w7} and the dashed curve represents the \emph{extremal} Reissner-Nordstr\"om black hole. For $Q=0$ the plots shows the relation $r_{\text{ext}}=\sqrt{4.567}\,\theta=2.137\theta$ as found in~\cite{SS2}.}}\label{Fig3}
\end{figure*}
\begin{figure}
\centering
\includegraphics[width=0.43\textwidth]{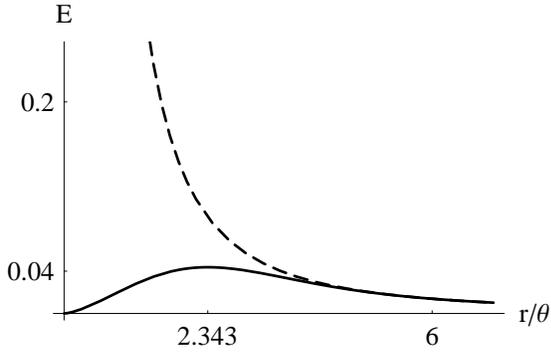} \\
\caption{\footnotesize{The electric field for $Q/(4\pi\epsilon_0\theta^2)=0.5$. Continuous plot: The ``quantum" electric field $E$~\eqref{e2} of the noncommutative black hole/charged quantum particle~\eqref{w7}. Dashed plot: The electric field of the Reissner-Nordstr\"om black hole.}}\label{Fig4}
\end{figure}
\begin{figure}
\centering
\includegraphics[width=0.43\textwidth]{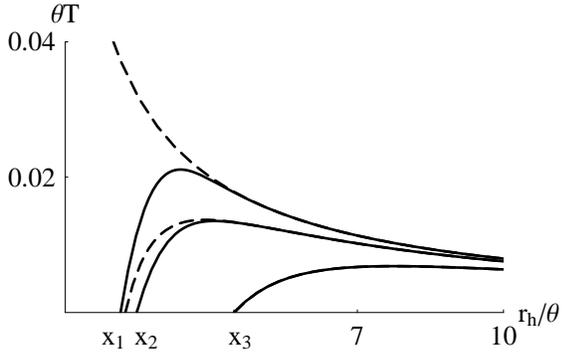} \\
\caption{\footnotesize{The temperature $T$ times the width $\theta$ [in units of $\hbar c/k_{\text{B}}$] versus $x\equiv r_h/\theta$ for $Q^2G/(4\pi\epsilon_0c^4\theta^2)=0,\,5,\,20$ from up to down. Continuous plots: The noncommutative charged black hole~\eqref{w7}. The values $x_1=2.137$, $x_2=2.470$, $x_3=4.465$ are the ratios $r_{\text{ext}}/\theta$ for the values $Q^2G/(4\pi\epsilon_0c^4\theta^2)=0,\,5,\,20$, respectively. Dashed plots: The Reissner-Nordstr\"om black hole. The plots for the noncommutative charged black hole and the Reissner-Nordstr\"om black hole almost coincide for $Q^2G/(4\pi\epsilon_0c^4\theta^2)=20$.}}\label{Fig5}
\end{figure}

The extremality condition corresponds to $f=0$ and $\partial_r f=0$ or, equivalently, to
\begin{equation}\label{w10}
(r^2f)\big|_{r_{\text{ext}}}=0\quad\text{ and }\quad \partial_r (r^2f)\big|_{r_{\text{ext}}}=0.
\end{equation}
Using the approach of Ref.~\cite{generate}, we solve these two equations for $M$ and $Q^2$ in terms of the extremal radius $r_{\text{ext}}$ then we parametric-plot $Q^2/\theta^2$ versus $M/\theta$ and $Q^2/\theta^2$ versus $M^2/\theta^2$ as in figures~\ref{Fig2} and~\ref{Fig3}, respectively. Figure~\ref{Fig2} includes a plot of the constraint~\eqref{w9} ensuring a de Sitter behavior in the vicinity of the origin: the region beneath the dashed straight line, which sets a limit for the stability of the solution, satisfies the constraint~\eqref{w9}. We see that stable charged quantum particles exist for $1.346\theta<M<4.254\theta$ and have $5.729\theta^2<Q^2<18.102\theta^2$ ($2.394\theta<|Q|<4.255\theta$). The same plot shows that black holes with $M>4.254\theta$ are stable if $Q^2/\theta^2 < \frac{16}{3}\sqrt{\frac{2}{\pi}}~M/\theta$ and unstable if $Q^2/\theta^2 > \frac{16}{3}\sqrt{\frac{2}{\pi}}~M/\theta$, while black holes with $M<4.254\theta$ are always stable (provided they are nonextremal).

The plot (a) of Fig.~\ref{Fig3} shows that for the same total mass parameter $M$, an extremal noncommutative charged black hole~\eqref{w7} is less charged than its Reissner-Nordstr\"om counterpart. By the plot (b) of Fig.~\ref{Fig3} we see that for the same mass $M$, an extremal noncommutative charged black hole~\eqref{w7} may have an extremal radius $r_{\text{ext}}$ greater (for relatively smaller values of $M$) or smaller (for relatively larger values of $M$) than that of its Reissner-Nordstr\"om counterpart.

\subsection{The electric field}
The electric charge density squared~\eqref{w6} being proportional to $r\e^{-r^2/(2\theta^2)}$, the corresponding electric charge density is then proportional to $\sqrt{r}\e^{-r^2/(4\theta^2)}$. We need a substitution rule, similar to~\eqref{i7} and~\eqref{i8}, for the electric charge density. This is of the form
\begin{equation*}
\frac{\delta(r)}{2\pi r^2}\to c\sqrt{r}\e^{-r^2/(4\theta^2)},
\end{equation*}
where we determine the constant $c$, as we did in~\eqref{i8}, by requiring
\begin{equation*}
\int _0^{\infty }c \sqrt{r}\e^{-r^2/(4\theta^2)} 4 \pi  r^2 dr=1.
\end{equation*}
This yields
\begin{equation}\label{i9}
\frac{\delta(r)}{2\pi r^2}\to \frac{\sqrt{r}\e^{-r^2/(4\theta^2)}}{16\pi\sqrt{2}\theta^{7/2}\Gamma(7/4)}.
\end{equation}
This way we obtain the ``quantum" charge density
\begin{equation}\label{e1}
\rho_q=\frac{Q\sqrt{r}\e^{-r^2/(4\theta^2)}}{16\pi\sqrt{2}\theta^{7/2}\Gamma(7/4)}.
\end{equation}
Notice that $\rho_q$ has a maximum value, as $\rho_{q^2}$ does, at $r=\theta$.

We can now proceed to the evaluation of the ``quantum" electric field $E$ by applying the Gauss Theorem $4\pi r^2E=(\int_0^{r}\rho_q(r')4\pi r'^2\dd r')/\epsilon_0$:
\begin{align}\label{e2}
E=&\frac{1}{4\pi\epsilon_0\Gamma(7/4)}\int_0^{r^2/(4\theta^2)}t^{3/4}\e^{-t}\dd t,\nn\\
\quad =&\frac{Q \big[\Gamma \big(\frac{7}{4}\big)-\Gamma \big(\frac{7}{4},\frac{r^2}{4 \theta ^2}\big)\big]}{4 \pi  \epsilon _0 \Gamma \big(\frac{7}{4}\big)r^2 }=\frac{Q
\gamma \big(\frac{7}{4},\frac{r^2}{4 \theta ^2}\big)}{4 \pi  \epsilon _0 \Gamma \big(\frac{7}{4}\big)r^2 },
\end{align}
[where we have set $t=r'^2/(4\theta^2)$] and this approaches the Reissner-Nordstr\"om value, $Q/(4 \pi  \epsilon _0 r^2)$, as $\theta\to 0$. Notice that $E$, which is brought to the form:
\begin{equation*}
E=\frac{Q}{4 \pi  \epsilon _0r^2 }-\frac{Q\Gamma \big(\frac{7}{4},\frac{r^2}{4 \theta ^2}\big)}{4 \pi  \epsilon _0 \Gamma \big(\frac{7}{4}\big)r^2 },
\end{equation*}
is always smaller than its Reissner-Nordstr\"om counterpart value as shown in Fig.~\ref{Fig4}. The electric field has its maximum value at $r=2.343\theta$ for all values of $Q$; this is the point where
$\gamma \big(\frac{7}{4},\frac{r^2}{4 \theta ^2}\big)/r^2$ has its maximum value.

The procedure given here is equivalent to directly solving $F^{\mu\nu}{}_{;\nu}\equiv (\sqrt{|g|}F^{\mu\nu}){}_{,\nu}/\sqrt{|g|}=J^{\mu}$ where $J^t$, given by~\eqref{e1}, is the only nonvanishing component of $J^{\mu}$.

\subsection{The temperature}
It is straightforward to evaluate the temperature,
\begin{equation*}
T=\frac{\hbar c}{4\pi k_{\text{B}}}~\partial_r f\big|_{r_{h}},
\end{equation*}
(where $\hbar$ and $k_{\text{B}}$ are the reduced Planck and Boltzmann constants)
of the noncommutative charged black hole in terms of $x\equiv r_h/\theta$ ($r_h$ is the radius of the event horizon) and $Q^2/\theta^2$,
\begin{align}\label{t1}
\frac{k_{\text{B}}}{\hbar c}\theta  T=&\frac{1}{4 \pi  x} \bigg(1-\frac{x^3\e^{-x^2/2}}{\sqrt{2} \gamma \big(\frac{3}{2},\frac{x^2}{2}\big)}\bigg)-\frac{\frac{Q^2G}{4 \pi  \epsilon _0 c^4 \theta ^2}}{4 \pi  x^3}\nn\\
&+\frac{\frac{Q^2 G}{4 \pi  \epsilon _0 c^4 \theta ^2}}{8 \pi  x^3} \bigg(2+x^2+x^4-\frac{\sqrt{2}
x^3}{\gamma \big(\frac{3}{2},\frac{x^2}{2}\big)}\bigg) \e^{-x^2/2}\nn\\
&+\frac{\frac{Q^2 G}{4 \pi  \epsilon _0 c^4 \theta ^2}}{8 \sqrt{2}
\pi  \gamma \big(\frac{3}{2},\frac{x^2}{2}\big)} (2+x^2) \e^{-x^2},
\end{align}
where we have eliminated the mass $M$ using $f(r_h)=0$. Its limit, as $\theta\to 0$, reduces to the value of the temperature for the Reissner-Nordstr\"om black hole,
\begin{equation*}
    \lim_{\theta\to 0}T=\frac{\hbar c}{4\pi k_{\text{B}}}~\frac{r_h^2-\frac{Q^2 G}{4 \pi  \epsilon _0 c^4}}{r_h^3}.
\end{equation*}
If we assume that the evaporation of the black hole via the Hawking radiation process carries only mass and no electric charge (no charged particles are emitted), we can investigate the process by plotting the temperature versus the radius of the event horizon for fixed $Q^2$. Figure~\ref{Fig5} is a plot of $\theta T$ for both the noncommutative charged black hole and the Reissner-Nordstr\"om black hole. The temperature is always finite for the noncommutative charged black hole for all values of $Q^2\geq 0$. The temperature has always a maximum value, $T_{\text{max}}$, that is monotonically increasing with $Q^2$. From Fig.~\ref{Fig5} we see that $T_{\text{max}}$ decreases with increasing $Q^2$. Since the back-reaction effects of the evaporation process, which are important as $T$ approaches $T_{\text{max}}$, are negligible for $Q^2=0$~\cite{SS2}, they are so for increasing values of $Q^2$.

By Fig.~\ref{Fig5} the evaporation process ends at $T=0$, the temperature of the extremal black hole. During the evaporation process, the black hole loses mass and energy until its remaining mass and its electric charge satisfy the extremality conditions~\eqref{w10} by the second of which $T=0$. However, the constraint of stability~\eqref{w9} may end the evaporation process at some nonzero temperature corresponding to
\begin{equation}\label{t2}
\frac{Q^2 G}{4 \pi  \epsilon _0c^4 \theta^2} \geq \frac{16}{3}\sqrt{\frac{2}{\pi}}~\frac{MG}{c^2\theta},
\end{equation}
which is just the condition contradicting~\eqref{w9}.

The emission of charged particles is theoretically possible in empty space via pair creation; we assume that it is also possible in a massive continuous medium (recall that the space around the noncommutative black hole is not empty).  If all particles created this way fall into the hole, its total charge remains unchanged. There are no available statistics on how many charges will fall in/escape from the hole. If the process went random, it would yield no change in the total charge of the black hole.

\begin{table*}
{\footnotesize
\begin{tabular}{|l|c|c|c|}
  \hline
  &  &  & \vspace{-0.3cm} \\
  {source} & $\epsilon$ & $p_r$ & $p_t$ \\
  &  &  & \vspace{-0.3cm} \\
  \hline
  \hline
  dust & $c^2 \rho _m$ & $0$ & $0$ \\
  \hline
  electric field & $\epsilon_E$ & $-\epsilon_E$ & $\epsilon_E$  \\
  \hline
  vacuum & $\frac{q^2(r)}{32 \pi ^2 \epsilon _0 r^4}-\epsilon_E-\frac{\rho _{q^2}}{8 \pi  \epsilon _0 r}$ & $-c^2 \rho _m+\frac{\rho _{q^2}}{8 \pi  \epsilon _0 r}+\epsilon_E-\frac{q^2(r)}{32 \pi ^2 \epsilon _0 r^4}$ & $-c^2 \rho _m-\frac{1}{2} c^2 r \partial_r\rho _m+\frac{\partial_r\rho _{q^2}}{16 \pi  \epsilon _0}+\frac{q^2(r)}{32 \pi ^2 \epsilon _0 r^4}-\epsilon_E$  \\
  \hline
\end{tabular}}
\caption{{\footnotesize The three components of the SET as dust, electric field, and vacuum. Here $\epsilon_E$ is the energy density $\epsilon_E$ of the electric field given by~\eqref{set1}, $\rho_m$ is given by~\eqref{i1}, and ($q^2(r),\,\rho _{q^2}$) are given by Eqs.~\eqref{w4}, \eqref{w5}, and~\eqref{w6}.}}\label{Tab1}
\end{table*}
\begin{table*}
{\footnotesize
\begin{tabular}{|l|c|c|c|}
  \hline
  &  &  & \vspace{-0.3cm} \\
  {source} & $\epsilon$ & $p_r$ & $p_t$ \\
  &  &  & \vspace{-0.3cm} \\
  \hline
  \hline
  anisotropic perfect fluid & $c^2 \rho _m$ & $\frac{\rho _{q^2}}{8 \pi  \epsilon _0 r}$ & $0$ \\
  \hline
  electric field & $\epsilon_E$ & $-\epsilon_E$ & $\epsilon_E$  \\
  \hline
  vacuum & $\frac{q^2(r)}{32 \pi ^2 \epsilon _0 r^4}-\epsilon_E-\frac{\rho _{q^2}}{8 \pi  \epsilon _0 r}$ & $-c^2 \rho _m+\epsilon_E-\frac{q^2(r)}{32 \pi ^2 \epsilon _0 r^4}$ & $-c^2 \rho _m-\frac{1}{2} c^2 r \partial_r\rho _m+\frac{\partial_r\rho _{q^2}}{16 \pi  \epsilon _0}+\frac{q^2(r)}{32 \pi ^2 \epsilon _0 r^4}-\epsilon_E$  \\
  \hline
\end{tabular}}
\caption{{\footnotesize The three components of the SET as anisotropic perfect fluid (with a positive radial pressure), electric field, and vacuum. Here $\epsilon_E$ is the energy density $\epsilon_E$ of the electric field given by~\eqref{set1}, $\rho_m$ is given by~\eqref{i1}, and ($q^2(r),\,\rho _{q^2}$) are given by Eqs.~\eqref{w4}, \eqref{w5}, and~\eqref{w6}.}}\label{Tab2}
\end{table*}

\subsection{The stress-energy tensor}
In the uncharged case~\eqref{i6} we have seen that the vacuum acquires negative radial and tangential pressures. The ``response" of the vacuum to the mass distribution by generating negative pressures proportional to the mass density cannot be understood within the approach of the phenomenological model presented here. This model stipulates that the effects of commutativity can, for distances larger than the Planck length, be appropriately incorporated as an extended, distributional, source terms in the field equations. The response of the vacuum is necessary to maintain a distributional source by exerting a repulsive force on the collapsing matter.

In the charged case, the response of the vacuum is linear to each source term in the field equations. Table~\ref{Tab1} provides the values of the energy density, radial and tangential pressures for the three components of the distributional source term: dust, electric field, and vacuum. The energy density $\epsilon_E$ of the electric field is given by
\begin{equation}\label{set1}
\epsilon_E=\frac{\epsilon_0E^2}{2}=\frac{\epsilon_0}{2}\bigg(\frac{Q
\gamma \big(\frac{7}{4},\frac{r^2}{4 \theta ^2}\big)}{4 \pi  \epsilon _0 \Gamma \big(\frac{7}{4}\big)r^2}\bigg)^2.
\end{equation}
It goes without saying that, for physical values of ($M,\,Q,\,\theta$), $p_r<0$ and $p_t<0$ for vacuum (fourth line in Table~\ref{Tab1}).

For the charged case, the expressions of the pressures for the vacuum include extra terms with respect to~\eqref{i6}, among which the term
\begin{equation}\label{set2}
\epsilon_E-\frac{q^2(r)}{32 \pi ^2 \epsilon _0 r^4}<0,
\end{equation}
[$q^2(r)$ is given by Eqs.~\eqref{w4} and \eqref{w5}] which is a quantum correction stating that the charge is no longer localized in a point. Said otherwise, the substitution rules~\eqref{i8} and~\eqref{i9} yield $q^2(r)\neq (q(r))^2$ and $Q^2=(Q)^2$. The correction term~\eqref{set2} and $q^2(r)- (q(r))^2$ tend to $0$ in both limits $\theta\to 0$ and $r\to \infty$. These substitution rules constitute new ways of counting.

The presence of an electric charge may cause the dust to acquire a radial pressure, which we have neglected in Table~\ref{Tab1}. One may split the SET as anisotropic perfect fluid (with a positive radial pressure), electric field, and vacuum as shown in Table~\ref{Tab2}. The distributions of the roles in Tables~\ref{Tab1} and~\ref{Tab2} are tentative ways to split the components of the SET into three physical sub-SETs; only the fourth line remains unjustified quantum mechanically. This way of splitting the SET is in any way better than assuming it to be that of a single anisotropic perfect fluid~\cite{SS2,SS3}.

\section{Conclusion\label{secc}}
This paper is a further step in the understanding of the physics of quantum black holes. We have presented a semi-classical model for a geometric description of quantum charged particles and noncommutative charged black holes that are asymptotically flat with a de Sitter behavior in the vicinity of the origin. This phenomenological description has been achieved upon using a metric that is a solution to the classical field equations with extended Gaussian mass, Weibull charge, and vacuum one-parameter distributions. The common width of the distributions could be identified with the Compton wavelength~\cite{SS3} inversely proportional to the mass of the extended source, yielding a variable parameter. As the mass increases the width tends to zero and the two extended distributions each become a point-like Dirac distribution resulting in a Reissner-Nordstr\"om black hole solution.

Do the solutions derived in~\cite{SS2,SS3,SS4} and in the present work are just ``arbitrary mathematical solutions"? Such a description applies to some wormhole and cosmological solutions derived upon running the field equations from left to right~\cite{Synge,Ellis}. In this work we have presented an almost direct derivation: We have fixed a Gaussian mass density based on a conclusion drawn in~\cite{SS2} and a Weibull charge density squared based upon the requirement that the solution should have a de Sitter behavior in the vicinity of the origin. We have then run the field equations from left to right and from right to left to determine the metric and the electric field.

We have shown that a quantum description of the black hole physics and geometry requires a set of substitution rules by which the point Dirac distribution is replaced by a one-parameter, smooth, family of distributions constituting new ways of counting. As the parameter tends to zero, the quantum properties of the black hole/quatum particle regain their classical values.

The difficulty in using a Gaussian distribution and its Weibull family is that the physical/geometrical properties cannot be given in closed forms. Moreover, the use of the Gaussian distribution has been inspired, but not fully justified, from the expression of the scalar product of two mean positions. In our subsequent paper we will introduce new mathematical one-parameter distributions, mimicking the Gaussian and Weibull distributions and reduce to Dirac distribution in the limit of vanishing parameter, yielding simple closed forms for all physical/geometrical entities.


\renewcommand{\theequation}{A.\arabic{equation}}
\setcounter{equation}{0}



\end{document}